# First Jump of Microgel: Actuation Speed Enhancement by Elastic Instability

Howon Lee, Chunguang Xia and Nicholas X. Fang*

**Swelling-induced snap-buckling in a 3D micro hydrogel device, inspired by the insect-trapping action of Venus flytrap, makes it possible to generate astonishingly fast actuation. We demonstrate that elastic energy is effectively stored and quickly released from the device by incorporating elastic instability. Utilizing its rapid actuation speed, the device can even jump by itself upon wetting.**

Adaptive materials that can sense and react to external signals have been attracting growing attention in various fields of science and engineering because of their potential for autonomous and multifunctional devices and systems. Hydrogels, which swell and contract in response to a wide range of environmental changes, have been intensively studied as one of the most promising functional materials.[1-7] However, the operation speed of hydrogel-based devices is inherently limited by the slow diffusion of solvent into polymer network.[2] On the other hand, nature offers a novel solution to breach this barrier as demonstrated in the insect-trapping action of Venus flytrap. This astonishingly fast motion is attained by involving snap-buckling instability.[8,9] Inspired by this exquisite mechanism, here we present rapid actuation of a micro hydrogel device by exploiting swelling-induced snap-buckling. Utilizing its fast actuation speed, the device can even jump upon wetting. We demonstrate that elastic energy is effectively stored and quickly released from the device by incorporating elastic instability. In our experiment, the micro device could generate a snapping motion within 12 milliseconds, releasing power at a rate of 34.2 mW g$^{-1}$. This engineered microgel will open up new gateways for intelligent systems in diverse fields, such as microfluidics, soft robotics, artificial muscle, and biomedical engineering.

Hydrogels are a network of polymer chains that undergoes volumetric change upon solvent absorption. Since the first demonstration of reversible expansion and contraction of hydrogels,[1] this unique phenomenon has been intensively studied experimentally[1-7] and theoretically.[10,11] More recently, many hydrogels have been found responsive to a variety of environmental stimuli such as temperature, pH, light, electric field and many others.[1-7] Along with the direct conversion of chemical energy to mechanical work, the responsive properties of hydrogels make it possible to integrate multiple functions including sensing and actuation in a single device, which leads to diverse applications in microfluidics,[2] optics,[3] and drug delivery.[12]

What remains as a challenge, however, is to fulfill a high speed operation in hydrogel devices. Swelling of hydrogel occurs when solvent diffuses into polymer network. Since diffusion time of solvent molecules scales with diffusion length of polymer as $L^2$ in general,[10] the need of fast actuation is pushing down the ultimate dimension of the hydrogel devices.[2] However, miniaturization often causes inevitable sacrifices in operating stroke, mechanical strength, and/or reliability. More importantly, required actuation speed of a device is not necessarily matched with dimension. Hence, the dependence of operation speed on dimension is definitely a hindrance in exploring further possibilities for hydrogel devices.

Interestingly, one can find a similar dilemma in plant kingdom. Plants developed unique movements for different purposes such as nutrition,[8] pollination,[13] and defence.[14] These elaborate motions are based on swelling of cells induced by internal fluid flow. The time scale of this hydro-elastic motion, called poroelastic time, is dependent on the square of the characteristic length scale of plants,[9] which is exactly the same limitation as found in hydrogel swelling. Surprisingly, though, some carnivorous and sensitive plants learned to make rapid motions regardless of their size using elastic instability.[8,9] In this mechanism, length scale is no longer a limiting factor. The Venus flytrap, for example, closes their leaves in a fraction of a second to trap insects. The doubly curved shape of their leaves leads to snap-buckling instability, making the leaf closure much faster beyond the poroelastic time.[8] In this way, Venus flytraps can have large but fast enough leaves to catch preys, which would not be possible without elastic instability.

Following nature's guide as implemented in Venus flytraps, we designed and fabricated a hydrogel device in a doubly curved shape to incorporate elastic instability (Fig. 1a). Projection micro-stereolithography (PμSL)[15] was used to fabricate a three dimensional (3D) micro device (see Experimental). The leaf is a 1200 μm by 1200 μm plate with thickness of 100 μm, and is convex outward with a radius of curvature of 800 μm. For solvent delivery, three microfluidic channels of 150 μm by 50 μm in cross-section are embedded with 400 μm spacing on the inner surface of the leaf. Thus, the device changes the curvature from convex to concave upon swelling. The embedded microfluidic channels play a critical role in generating a 3D motion. Unlike other hydrogel systems in which the entire device needs to be immersed in solvent for actuation, swelling in our device takes place locally around the channels by direct solvent delivery. Therefore, a desired motion can be obtained by embedding channels where the swelling is needed. In addition, wedge-shaped cross-section of the channel (Fig. 1b) not only promotes simple and fast delivery of solvent via capillary force, but also allows fast evaporation of solvent through the opening for quick geometrical restitution when the device is allowed to dry.

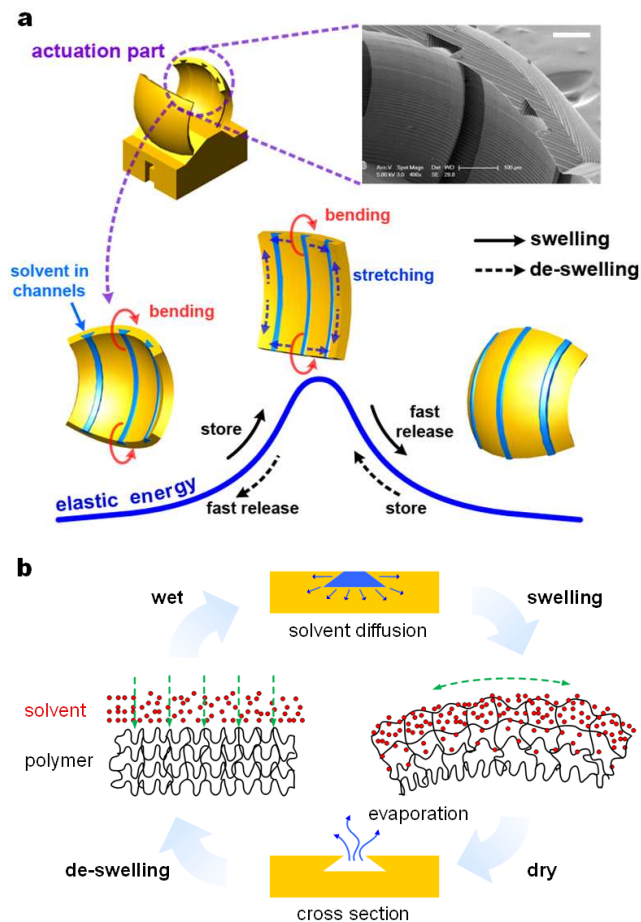

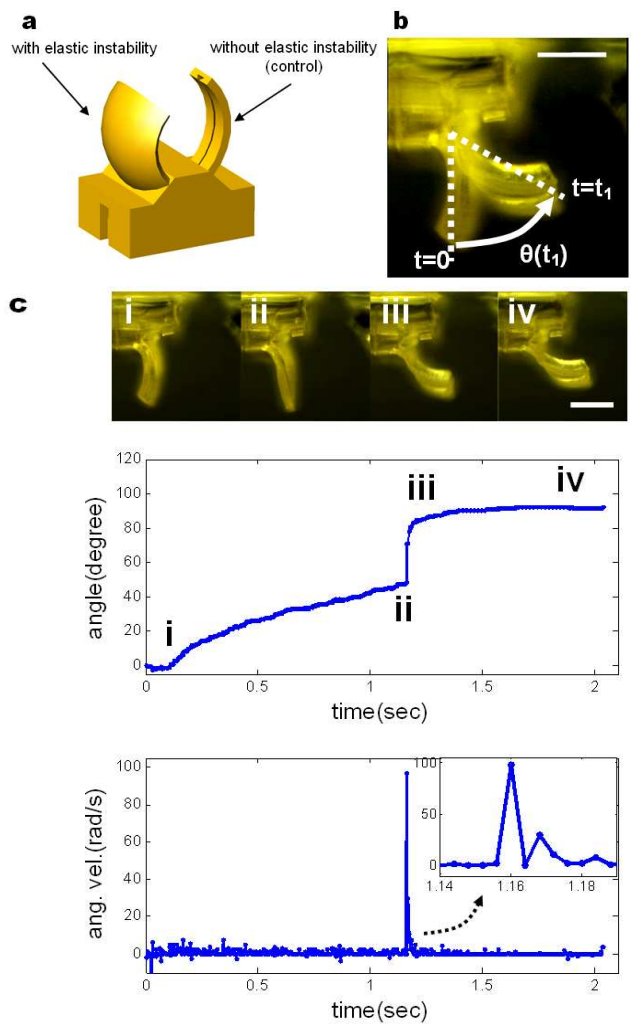

**Fig. 1** Swelling-induced snap-buckling in a doubly curved hydrogel device with embedded microfluidic channels. (a) 3D micro hydrogel device in doubly curved shape and scanning electron microscope (SEM) image of embedded microfluidic channels. Scale bar indicates 100 μm. (b) Swelling and de-swelling mechanism using embedded microfluidic channels.

For snap-buckling instability, it is essential for the device to have directional swelling capability in addition to the doubly curved shape. The Venus flytrap, for instance, actively regulates internal hydraulic pressure to bend their leaf in the direction perpendicular to the midlib.[8] In mechanical point of view, the leaf is an elastic plate in doubly curved shape. When bent along one axis only, elastic body becomes stretched as a result of bending-stretching coupling of doubly curved plate, thereby storing elastic energy. As the leaf further deforms and passes through the energy barrier, stored elastic energy is instantaneously released and converted into kinetic energy, creating a rapid trap closure (see Electronic Supplementary Information, Fig. S4). Likewise, the swelling of the device needs to be controlled in such a way that the curvature is actively manipulated only in one direction while the curvature in the other direction remains passive. Although there have been a few reports of fast polymer actuation using elastic instability,[16, 17] accurate control of the shape has not been demonstrated so far due to the lack of local swelling control. To achieve directional swelling of doubly curved hydrogel device, we aligned the microfluidic channels vertically with spacing in between. Once supplied to the channels, solvent diffuses out in all directions around the channel. However, if the distance between the channels is far enough (400 μm), swelling in lateral direction is negligible within a short period of actuation time since the solvent diffusion speed is finite. On the contrary, swelling along the channels can quickly cause significant amount of bending in vertical direction as the diffusion length in thickness direction is relatively short (50 μm). Therefore, local swelling around the aligned channels makes the doubly curved device bend only along the vertical axis and, as a result, snap-buckling occurs. Similarly, the device snaps back to the original shape during de-swelling. This process is illustrated in Fig. 1a.

**Fig. 2** Enhancement of actuation speed by elastic instability. (a) 3D micro hydrogel device with elastic instability and a control sample. (b) Angle measurement during actuation. The doubly curved leaf points downwards. (c) The shape change of the device during actuation (see Electronic Supplementary Information, Movie 1). Snap-buckling occurs between **ii** and **iii**. All scale bars indicate 500 μm.

To investigate the effect of elastic instability on motion speed, we captured the motion of the device using high speed camera (Redlake Image, 250 fps). Image analysis software is used to measure the angle of the line connecting the tip and the base of the device from the initial configuration at each frame (Fig. 2b). Geometric change in the snapping process of the doubly curved plate is shown in Fig. 2c. The angle changed abruptly from 48 ° to 80 ° in 12 milliseconds with a maximum angular velocity as high as 100 rad s$^{-1}$ (see Electronic Supplementary Information, Movie 1). This is an exceptionally high speed to obtain in a hydrogel device. In comparison with natural system, angular velocity observed in most Venus flytraps' leaves is below 10 rad s$^{-1}$.[8] As a control experiment, we tested another sample that has the same curvature in vertical direction but has no curvature (flat) in horizontal direction (Fig. 2a). The control sample therefore experiences just uniaxial bending without elastic instability during actuation. The width of the control sample (400 μm) is one third of the doubly curved sample and it has a single channel in the middle, thus its structural resistance against swelling actuation is the same as the doubly curved sample. Resulting motion of the control sample exhibits a smooth and continuous angle change with a maximum angular velocity less than 10 rad s$^{-1}$ (see Electronic Supplementary Information, Fig. S5 and Movie 2). The sharp contrast of these two experiments shows remarkable enhancement of actuation speed of hydrogel device via elastic instability.

To demonstrate the maximum use of enhanced speed obtained by swelling-induced snap-buckling, we made a micro hydrogel device jump upon wetting (Fig. 3). In this sample, both legs of the device are doubly curved hydrogel plates with embedded microfluidic channels. Hence, they open and close rapidly via snap-buckling during swelling and de-swelling, respectively. When a solvent droplet is applied, swelling causes the legs to open outward and consequently the device sits on the ground. When the device returns to its original configuration as the solvent evaporates, the legs close quickly by snap-buckling, generating a thrust to make the device itself jump off the ground (see Electronic Supplementary Information, Movie 3). It flew about 7 mm, which is about five times as long as its own length (see Electronic Supplementary Information, Fig. S7). As the elastic energy stored in the device is converted into kinetic energy during snap-buckling, we can estimate how much energy was stored and released from the device by measuring the jumping trajectory. With the actuation time of 12 milliseconds, power density of the device was estimated at 34 mW g$^{-1}$ (see Experimental), which is comparable to that of our own muscle (100 mW g$^{-1}$).[18]

We presented a 3D microgel actuator and a method to significantly enhance the actuation speed of the devices by incorporating elastic instability. This actuation strategy can be also associated with a wide range of external stimuli (temperature, pH, light, etc.) with a proper choice of stimuli-responsive hydrogel. We believe that adding agility to the integrated functionalities of hydrogel will greatly extend the potential of hydrogel devices and systems. The fast, responsive, and multifunctional devices enabled by this approach will allow us to envision entirely new opportunities

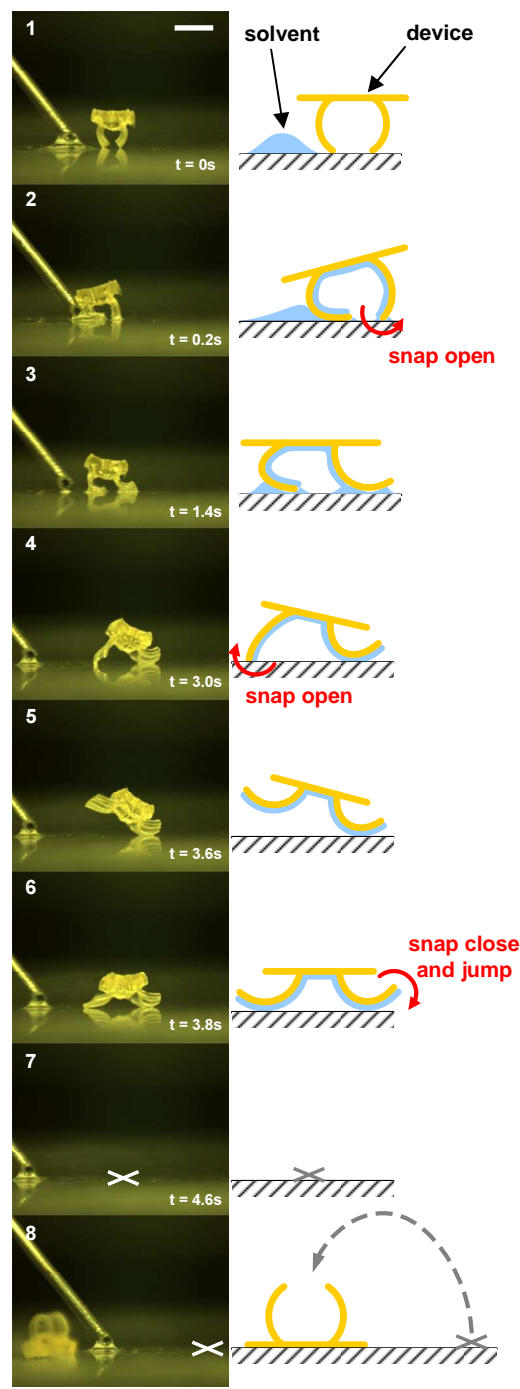

**Fig. 3** Microgel jumps upon wetting (see Electronic Supplementary Information, Movie 3). Numbers indicate the sequence of the motion. The device is placed on a glass substrate (1). When a solvent droplet is applied to the left leg (2), the solvent not only surrounds the leg, it also fills all the microfluidic network in the device by capillary action. Even after solvent surrounding the legs evaporates, embedded microfluidic channels are still filled with solvent, resulting in bending of legs outwards (3-5). With two legs bent outwards, the device is ready to jump (6). As the solvent in the channels further evaporates, snap-buckling takes place as the legs snap back to the original shape in de-swelling process. This rapid motion

produces enough thrust for jumping. The device jumps out of the view field (7) and is found outside the initial view field (8). Cross mark indicates the initial position of the device. Scale bar indicates 1 mm.

in various fields, such as self-operating pumps and fast-switching valves in microfluidic devices, soft robots with artificial muscle, and fast drug delivery vehicles for prompt measures in medical emergency.

## Experimental

**Materials**: The prepolymer solution used to make the hydrogel consists of poly(ethylene glycol) diacrylate (PEGDA) (MW~575, Sigma-Aldrich) with 2.0 wt.% photoinitiator (Irgacure 819, Ciba) and 0.75 wt.% photoabsorber (Sudan I, Sigma-Aldrich). The solvent used for swelling is acetone (99.8%, Sigma-Aldrich).

**3D microfabrication**: As a fabrication method for 3D polymeric device with embedded microfluidic channels, we used projection micro-stereolithography (PµSL) (see Electronic Supplementary Information, Fig. S1).[15] It is capable of fabricating intricate three dimensional micro structures in a layer-by-layer fashion with a resolution up to 2 µm. Firstly, 3D CAD drawing of the device is sliced into a set of layers. The image of each layer is sent to a Liquid Cristal on Silicon (LCoS, Canon) chip which plays a role as a dynamic mask generator. A flood of UV light (436 nm) from mercury lamp is reflected off the dynamic mask and then the beam containing the image is optically routed and focused on the surface of prepolymer solution through the 10x reduction projection lens in the light path. Although fabrication area for one exposure is about 1200 µm by 800 µm, larger area can be covered using x-y stages. Once a layer is polymerized on the substrate, the z-stage drops the substrate under the surface of the solution by the thickness of the next layer, and the machine projects the next image to polymerize the following layer on top of the preceding layer. This process proceeds iteratively until all the layers are completed. Layer thickness of the doubly curved plate and the base of the device in this study are 15 µm and 25 µm, respectively. Light intensity on the prepolymer surface was 6.17 mW cm$^{-2}$ and exposure time for 15 µm and 25 µm thick layers were 12 s and 20 s, respectively.

**Estimation of energy release**: From the jumping trajectory, we estimated energy release during snap-buckling. Overall, stored elastic energy is converted into energy in several forms such as translation energy, rotation energy, internal dissipation by viscosity, energy required to overcome the stiction to the ground when it jumps, and energy consumed by drag from the air.

$$E_{elastic} = E_{translation} + E_{rotation} + E_{dissipation} + E_{stiction} + E_{drag}$$

The parabolic trajectory of jumping (see Electronic Supplementary Information, Fig. S7) enables us to calculate $E_{translation}$ which can be an estimation of the minimum energy released from the snapping motion because $E_{elastic} > E_{translation}$. The measured distance and height of the parabolic trajectory are 7.33 mm and 2.11 mm, respectively. Provided that the trajectory is an ideal parabola following $(x, y) = (v_0 t \cos\theta, v_0 t \sin\theta - gt^2/2)$, we can calculate the launching angle $\theta = 49°$ and initial velocity $v_0 = 260.7$ mm/s. Thus, it gives

$$E_{translation} = \frac{1}{2}mv_0^2 = \frac{1}{2}(0.75 \text{ mg})(260.7 \text{ mm s}^{-1})^2 = 25.5 \text{ nJ}$$

where $m = 0.75$ mg is the total mass of the device in dry state. Energy density of this device can be estimated by dividing the energy by the mass of the actuation part (a leg) only. With the actuation time of 12 milliseconds, the power density of the device can be estimated as follows.

Energy density = $E_{translation} / m_{leg}$ = 25.5 nJ / 0.062 mg = 0.41 mJ g$^{-1}$

Power density = (0.41 mJ g$^{-1}$) / (12 ms) = 34.2 mW g$^{-1}$

## Acknowledgement

The authors would like to thank Professor N. G. Glumac for the permission to use the high speed camera.

## Notes and references

*Department of Mechanical Science and Engineering, University of Illinois at Urbana-Champaign, Urbana, IL, USA. Fax: 217-244-9956; Tel: 217-265-8262; E-mail: nicfang@illinois.edu*

† Electronic Supplementary Information (ESI) available: Details of fabrication, characterization, experiment and analysis are provided. Also included are videos of motions presented in Fig. 2 and Fig. 3. See DOI: 10.1039/b000000x/